\documentclass[superscriptaddress,aps,pra,singlecolumn]{revtex4}
\usepackage{amsmath,amssymb,graphicx}
\usepackage{color}

\newsavebox{\smlmat}
\savebox{\smlmat}{$\left(\begin{matrix}\cos{\theta}&\sin{\theta}\\\sin{\theta}&-\cos{\theta}\end{matrix}\right)$}

\begin{document}

\title{Deterministic Local Expansion of W States}

\author{Can Yesilyurt}
\thanks{Present address: Electrical and Computer Engineering, National University of Singapore, Singapore 117576, Republic of Singapore.}
\affiliation{Department of Computer Engineering, Okan University, Tuzla, Istanbul, 34959, Turkey}

\author{Sinan Bugu}
\affiliation{Department of Computer Engineering, Istanbul University, Beyazit, Istanbul, 34452, Turkey}

\author{Fatih Ozaydin}
\email{MansurSah@gmail.com}
\affiliation{Department of Information Technologies, Isik University, \c{S}ile, Istanbul, 34980, Turkey}

\author{Azmi Ali Altintas}
\affiliation{Department of Electrical Engineering, Okan University, Tuzla, Istanbul, 34959, Turkey}

\author{\\Mark Tame}
\affiliation{School of Chemistry and Physics, University of KwaZulu-Natal, Durban 4001, South Africa}
\affiliation{National Institute for Theoretical Physics, KwaZulu-Natal, South Africa}

\author{Lan Yang}
\affiliation{Department\,of\,Electrical\,and\,Systems\,Engineering, Washington University, St. Louis, MO 63130, USA}

\author{\c{S}ahin Kaya \"Ozdemir}
\affiliation{Department\,of\,Electrical\,and\,Systems\,Engineering, Washington University, St. Louis, MO 63130, USA}

\begin{abstract}
In large quantum systems multipartite entanglement can be found in many inequivalent classes under local operations and classical communication. Preparing states of arbitrary size in different classes is important for performing a wide range of quantum protocols. $W$ states, in particular, constitute a class with a variety of quantum networking protocols. However, all known schemes for preparing $W$ states are probabilistic, with resource requirements increasing at least sub-exponentially. We propose a deterministic scheme for preparing $W$ states that requires no prior entanglement and can be performed locally. We introduce an all-optical setup that can efficiently prepare $W$ states of arbitrary size. Our scheme advances the use of $W$ states in real-world quantum networks and could be extended to other physical systems.
\end{abstract}

\maketitle


\section{Introduction}

Entanglement is an elementary ingredient for realizing many quantum information processing tasks~\cite{Horodecki2009}, including quantum teleportation~\cite{Bennett1993}, superdense coding~\cite{Bennett1992}, quantum key distribution~\cite{Gisin2002}, quantum metrology~\cite{PezzeSmerzi2009PRL} and quantum computing~\cite{Steane1998}. Bipartite entanglement is well understood both theoretically and experimentally, whereas multipartite entanglement continues to experience intense effort from theorists and experimentalists. Here, the complexity of preparing, characterizing and manipulating various types of multipartite entangled states makes them challenging to study. A main difficulty in understanding entanglement in multipartite systems is that there is no unique type of multipartite entanglement, but rather different classes, such as GHZ~\cite{Greenberger1989}, $W$~\cite{Dur2001}, Dicke~\cite{Dicke1954} and cluster~\cite{Briegel2001}, and the majority of states from these classes cannot be converted easily into each other, for instance, using local operations and classical communication~\cite{Dur2000}.

Recently, interest has been steadily growing in the study of complex quantum network architectures for distributed quantum communication and computing. Such networks require multipartite entanglement shared among the various nodes. At the same time, there has also been a great deal of interest from researchers in finding specific entanglement classes that may be more efficient than others for carrying out a given task. For example, GHZ states can be used to reach consensus in distributed networks without classical post-processing; $W$ states have been proposed for leadership election in anonymous quantum networks~\cite{Hondt2006} and a number of secure quantum communication protocols~\cite{Joo2002,Wang2007,Cao2006,Liu2011}; cluster and graph states form the universal resources for quantum computing~\cite{Briegel2001}; and symmetric Dicke states have been shown to be useful resources for playing quantum versions of classical games~\cite{Ozdemir2007NJP}, and for symmetric telecloning and secret sharing~\cite{MarkPRL2009}. The generation of large-scale multipartite entangled states among many participants, or nodes of a network, is clearly a key step in carrying out such distributed quantum information processing tasks. Consequently, there has been much effort in proposing schemes to prepare and characterize multipartite entangled states in different physical systems, ranging from nuclear magnetic resonance~\cite{NMRRefs1,NMRRefs2,NMRRefs3,NMRRefs4}, cavity QED~\cite{CavityRefs1,CavityRefs2,CavityRefs3} and ion traps~\cite{IonTrapRefs1,IonTrapRefs2,IonTrapRefs3,IonTrapRefs4} to superconducting systems~\cite{SuperCondRefs1,SuperCondRefs2,SuperCondRefs3} and photonics~\cite{PhotonicsRefs1,PhotonicsRefs2}. Recent studies have also proposed and realized quantum gates for manipulating, expanding and fusing entangled states to form larger multipartite entangled states in quantum networks ~\cite{RecentW1,RecentW2,RecentW3}. The focus of research in this direction has mainly been on performing expansion or fusion operations by locally accessing only a limited number of qubits, while allowing for elements of classical communication.

Photonics in particular has been a popular and readily accessible test-bed for the study of different classes of entangled states. The experimental preparation and characterization of GHZ, $W$, Dicke and cluster states up to six-qubits have become routine in many quantum optics laboratories worldwide. For example, Kiesel {\it et al.} have demonstrated the preparation of a three-qubit $W$ state from a four-qubit symmetric Dicke state by projection of one of the qubits onto the computational basis~\cite{Kiesel2007PRL}. The initial Dicke state and the final $W$ state are suitable for specific tasks: the symmetric Dicke state can be used in quantum versions of classical games, whereas the $W$ state cannot~\cite{Shimamura2007PLA}. Instead, the $W$ state can be used for asymmetric telecloning, but the Dicke state cannot~\cite{Fan2014PhysRep}. Cluster states of different topologies have also been prepared in numerous experiments, and used for the realization of simple quantum algorithms and networking protocols~\cite{ClusterRefs1,ClusterRefs2,ClusterRefs3,ClusterRefs4,ClusterRefs5,ClusterRefs6}. In addition to all these experimental works, there have been many theoretical studies which have been aimed at finding optimal preparation and characterization approaches for multipartite entangled states. It is now known that GHZ and cluster states of arbitrary sizes can, in principle, be prepared deterministically, {\it i.e.} the schemes are limited only by experimental issues~\cite{Browne2005,Zeilinger1997}. However, this is not the case for $W$ states and Dicke states. The main difficulty for $W$ states and their generalisation in the form of Dicke states is that a deterministic transformation is impossible for given states via accessing only one qubit. The minimum number of qubits that need to be accessed and the limits of transforming one Dicke state to another Dicke state have been recently studied~\cite{Kobayashi2014}. There have also been theoretical proposals and some experimental implementations for creating $W$ states by expanding or fusing given $W$ states (or Bell states) with ancilla photons via accessing a single qubit of the given state and with a minimal gate complexity~\cite{Tashima2008,Tashima2009A,Tashima2009B,Ikuta2011}. In a recent experiment, Tashima {\it et al.} have shown the expansion of $W$ states using a very simple optical gate by accessing only one qubit of the $W$ state~\cite{Tashima2010}.

Due to the specific structure of entanglement in $W$ states, any transformation via accessing only a single qubit of the state -- regardless of the experimental imperfections -- leads to a probabilistic process~\cite{Ozdemir2011}. Therefore, the cost of preparing $W$ states (in terms of consumed Bell pairs) increases exponentially with respect to the target size. Recently, Ozdemir {\it et al.} have proposed fusing two $W$ states of arbitrary sizes (each larger than two qubits) to prepare larger $W$ states with a sub-exponential resource cost. This cost is due to a strategy that recycles the non-destroyed $W$ states in the probabilistic fusion process~\cite{Ozdemir2011}. The basic concept of the fusion gate was further developed by a series of theoretical proposals which enhanced the efficiency of the process~\cite{Bugu2013A}. Subsequently, a method to fuse multiple $W$ states simultaneously to reach a target size with a smaller number of fusion processes was proposed~\cite{Yesilyurt2013A,Ozaydin2014A}. Using more sophisticated gates and accessing more qubits of each given state, one would expect to increase the success probability of fusion and expansion operations. Despite a considerable amount of effort put into studying various aspects of $W$ states~\cite{WWorks1,WWorks2,WWorks3,WWorks4,WWorks5,WWorks6,WWorks7,WWorks8}, it is surprizing that, as of yet, there has been no scheme proposed that can deterministically prepare $W$ states with arbitrary size.

Here, we introduce a scheme for the deterministic preparation of $W$ states of arbitrary size and give an example optical setup for implementing the scheme using polarization-entangled $W$ states. In particular, the deterministic scheme can perform the following tasks: (i) prepare a polarization-entangled Einstein-Podolsky-Rosen (EPR) photon pair starting with two independent single photons, (ii) prepare a four-photon $W$ state from an EPR photon pair and two ancilla photons, and (iii) double the size of a $W$ state with the help of ancilla photons by running the scheme in parallel on each qubit of the $W$ state. In this final case, the scheme can be used to deterministically expand a $W$ state of $n$ qubits to a $W$ state of $2n$ qubits using $n$ ancilla qubits. Furthermore, it can probabilistically expand the size of a $W$ state if access to all qubits is not allowed, or by adding one-qubit at a time by repeatedly running the circuit on only one qubit of the state. Even in such extreme cases, the success probability of our scheme is much higher than all previously proposed schemes.
\begin{figure}[t]
\centerline{\includegraphics[width=0.5\columnwidth]{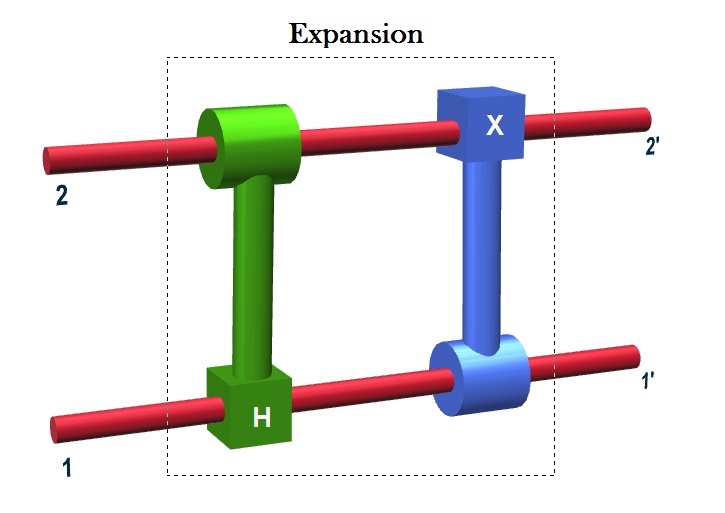}}\label{fig:scheme}
\caption{The proposed expansion scheme. A controlled-Hadamard (CH) gate with the control qubit in mode 2 and target qubit in mode 1 is followed by a controlled-NOT (CNOT) gate with the control qubit in mode 1 and target qubit in mode 2.}
\end{figure}



\section{A gate for expanding $W$ states and its cascaded operation}

The basic building block of the expansion circuit we propose is shown in Fig.~1. This circuit is composed of a controlled-Hadamard (CH) and a controlled-NOT (CNOT) gate. The CH gate can be decomposed into a CNOT gate and four single-qubit gates, as shown in Fig.~2~a. For an input state $|a\rangle_1|b\rangle_2$ in modes 1 and 2 with $\{a,b\}=\{0,1\}$, the CH gate, with the qubit in mode 2 as the control qubit, performs the transformation
\begin{eqnarray}\label{eq:1}
 |a\rangle_1|b\rangle_2 &\to& \frac{|c\rangle_{1'}+(-1)^{c\oplus1}b|c\oplus1\rangle_{1'}}{ (\sqrt{2})^b}|b\rangle_{2'},
 \end{eqnarray}
where $c=a\oplus b$. The action of the CNOT gate, with the first qubit as the control qubit, on an input state $|a\rangle_1|b\rangle_2$ is given by $|a\rangle_1|b\rangle_2 \longrightarrow |a\rangle_{1'}|a\oplus b\rangle_{2'}=|a\rangle_{1'}|c\rangle_{2'}$.
Thus, we can describe the action of the total circuit in Fig. 1 as
 \begin{eqnarray}\label{eq:2}
\hskip-0.4cm |a\rangle_1|b\rangle_2 &\to& \frac{|c\rangle_{1'}|a\rangle_{2'}+(-1)^{c\oplus1}b|c\oplus1\rangle_{1'}|a\oplus1\rangle_{2'}}{(\sqrt{2})^b}.
 \end{eqnarray}
 \begin{figure}[t]
\centerline{\includegraphics[width=0.75\columnwidth]{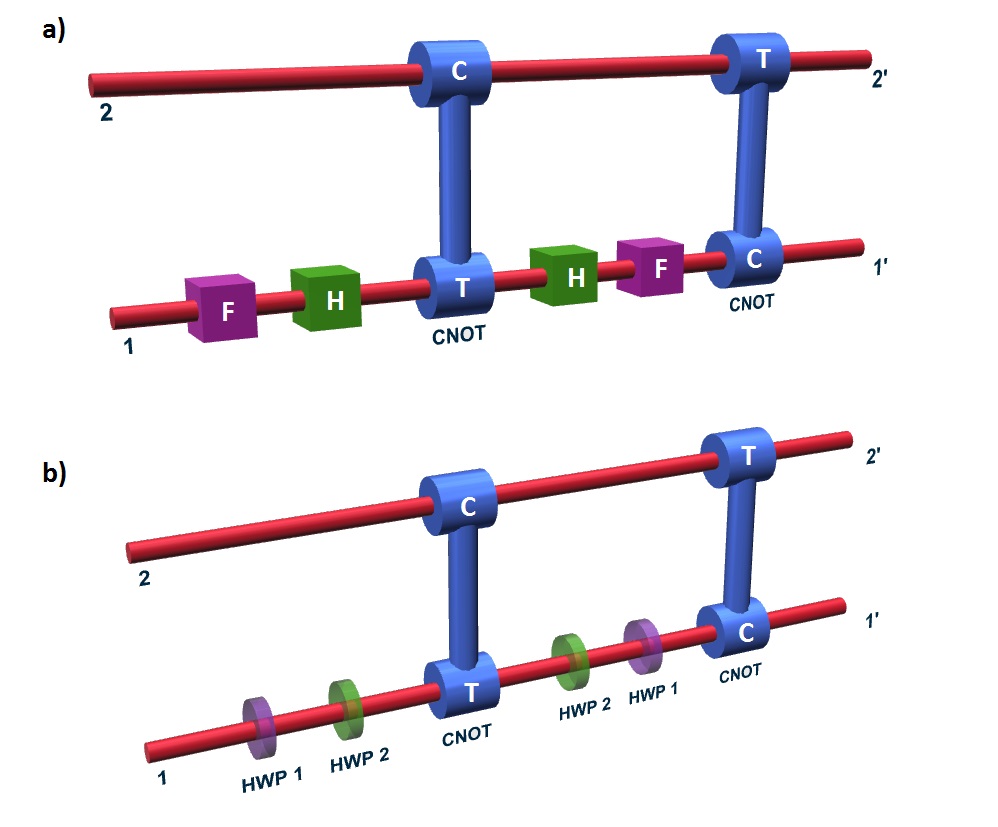}} \label{fig:optical}
\caption{Decomposition of the expansion circuit. {\bf (a)} Decomposition in the network model into two CNOT gates and four single-qubit gates. {\bf (b)} An optical setup for realizing the expansion circuit, where the $F$ and $H$ gates are implemented by half-wave plates (HWP) with the operation
$ HWP( \frac{\theta}{2})=$~\usebox{\smlmat} oriented at angles of $\pi /16$ and $\pi /8$, respectively~\cite{Miranowicz2010}. The CNOTs are implemented as outlined in Refs.~\cite{Brien2003,Nemoto2004,Brien2009,Okamoto2011}.}
\end{figure}

It is easy to see that the circuit prepares the Bell state $|\Psi^{+}\rangle_{1'2'}=(|01\rangle+|10\rangle)_{1'2'}/\sqrt{2}$ from the input $|01\rangle_{12}$, and the Bell state $|\Psi^{-}\rangle_{1'2'}=(|01\rangle-|10\rangle)_{1'2'}/\sqrt{2}$ from the input $|11\rangle_{12}$. For the other inputs, the transformation is as follows: $|00\rangle_{12}\to |00\rangle_{1'2'}$ and $|10\rangle_{12}\to |11\rangle_{1'2'}$, implying that for these inputs no entanglement is generated and thus the circuit in Fig. 1 is an entangling gate only if the inputs are properly chosen. Using the encoding $|0\rangle\equiv|H\rangle$ and $|1\rangle\equiv|V\rangle$, where $H$ ($V$) corresponds to the horizontal (vertical) polarization of a photon used to embody the qubit, this circuit prepares polarization entangled Bell states. In Fig.~2~b, we present the construction of the expansion circuit to process polarization encoded photonic qubits. In order to highlight this practical implementation of our scheme, for the rest of this work we will employ the polarization basis to represent the states.

Now let us analyze the action of the circuit when the photon in mode 1 is an ancilla photon $|H\rangle_{1}$ and the photon in mode 2 is provided from a polarization entangled Bell state $|\Psi^{+}\rangle_{20'}=(|HV\rangle_{20'}+|VH\rangle_{20'})/\sqrt{2}$ in modes 2 and $0'$, which can be assumed to be prepared using the same circuit as discussed above. The total `cascaded' circuit is shown in Fig.~3. The second expansion circuit in Fig.~3 performs the following transformation (where we have relabelled $2' \to 2$ from the output of the first expansion circuit for convenience)
\begin{eqnarray}\label{eq:3a}
&&\hskip-0.8cm |H\rangle_1|H\rangle_2|V\rangle_{0'} \to |H\rangle_{1'}|H\rangle_{2'}|V\rangle_{0'}\\
&&\hskip-0.8cm |H\rangle_1|V\rangle_2|H\rangle_{0'} \to \frac{|H\rangle_{1'}|V\rangle_{2'}|H\rangle_{0'}+|V\rangle_{1'}|H\rangle_{2'}|H\rangle_{0'}}{\sqrt{2}}.
 \end{eqnarray}
Subsequently we find
\begin{eqnarray}\label{eq:3}
|H\rangle_1|\Psi^{+}\rangle_{20'} \to \frac{\sqrt{2}}{2}(|HHV\rangle+|HVH\rangle+|VHH\rangle)_{1'2'0'}
 \end{eqnarray}
which corresponds to a weighted three-qubit $W$ state. The weights of this output state can be made equal using polarization dependent loss in mode $0'$ which performs the task $|H\rangle\rightarrow|H\rangle$ and $|V\rangle\rightarrow|V\rangle/\sqrt{2}$. As a result, the output state will be a three-qubit $W$ state of the form
\begin{eqnarray}\label{eq:3b}
 && \hskip-0.8cm |H\rangle_1|\Psi^{+}\rangle_{20'} \to \nonumber \\
 && \hskip-0.8cm\frac{\sqrt{3}}{2}\bigg(\frac{1}{\sqrt{3}}(|HHV\rangle+|HVH\rangle+|VHH\rangle)_{120'}\bigg) \nonumber \\
 && \hskip-0.8cm = \frac{\sqrt{3}}{2}|W_{3}\rangle_{1'2'0'},
 \end{eqnarray}
implying that the circuit prepares a genuine $W$ state with probability $p(W_2\rightarrow W_3)=3/4$. We will show in the next section how, instead of using a cascaded approach, `parallel' use of the expansion circuit enables deterministic expansion.
\begin{figure}[b]
\centerline{\includegraphics[width=0.7\columnwidth]{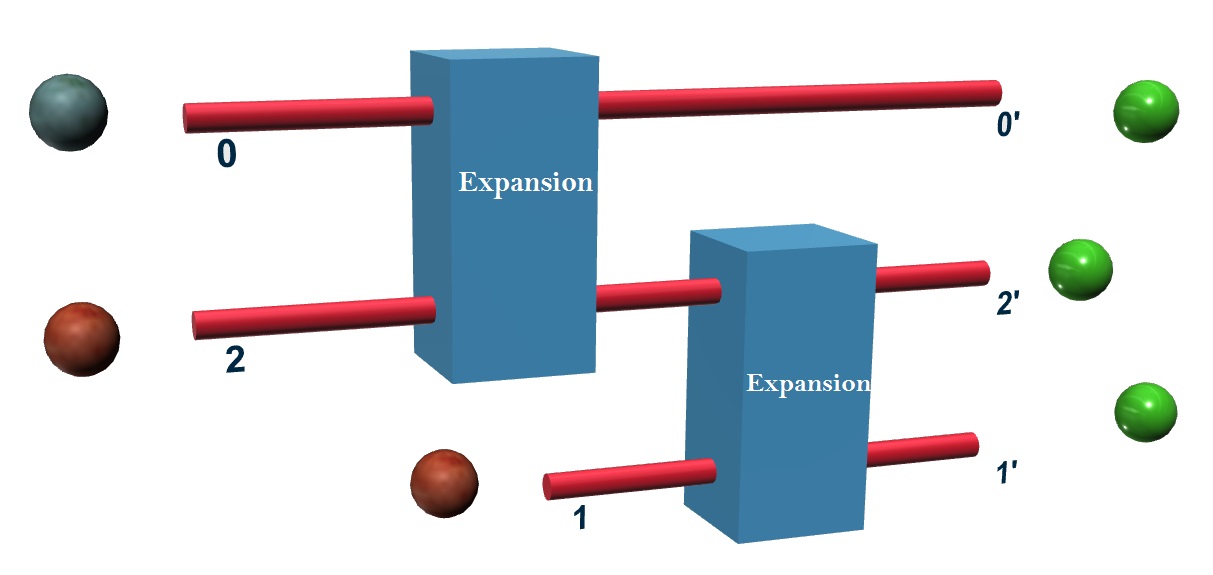}}\label{fig:cascaded}
\caption{$W$ state expansion starting from independent photons. Here the polarizations of the photons are $|V\rangle$ (blue, in mode 0), $|H\rangle$ (red, in mode 2) and $|H\rangle$ (red, in mode 1).}
\end{figure}

First, we discuss how cascaded use of the circuit can expand a general $W$ state of $N$ modes, $|W_{N} \rangle=|(N-1),1\rangle/\sqrt{N}$, where $|(N-1),1\rangle$ is the sum over all the terms with $N-1$ modes in $|H\rangle$ and one mode in $|V\rangle$. We can rewrite this state as $|W_{N} \rangle_{2 {\bf 0'}}=\big(|H\rangle_2 \otimes |N-2,1\rangle_{\bf 0'}+|V\rangle_2 \otimes |N-1,0\rangle_{\bf 0'}\big)/\sqrt{N}$, where mode 2 is separated out from the rest (which are denoted as ${\bf 0'}$ for convenience) and it will be input to mode 2 of the expansion circuit, while mode 1 of the circuit will have an ancilla photon $|H\rangle_1$, as before. The expansion circuit will perform the transformations
\begin{eqnarray}\label{eq:3c}
 && \hskip-0.6cm |H\rangle_1|H\rangle_2|N-2,1\rangle_{\bf 0'} \to |H\rangle_{1'}|H\rangle_{2'}|N-2,1\rangle_{\bf 0'},\\
 && \hskip-0.6cm |H\rangle_1|V\rangle_2|N-1,0\rangle_{\bf 0'} ~\to \nonumber \\
 && \hskip1.5cm \frac{[|H\rangle_{1'}|V\rangle_{2'}+|V\rangle_{1'}|H\rangle_{2'}]|N-1,0\rangle_{\bf 0'}}{\sqrt{2}}.
 \end{eqnarray}
Subsequently we find
\begin{eqnarray}\label{eq:3d}
 && \hskip-0.4cm |H\rangle_1|W_{N} \rangle_{2 {\bf 0'}} \to \nonumber \\
 && \hskip-0.4cm \frac{1}{\sqrt{N}}\bigg( |H\rangle_{1'}|H\rangle_{2'}|N-2,1\rangle_{\bf 0'} \nonumber \\
 && \hskip0.6cm+\frac{1}{\sqrt{2}} [|H\rangle_{1'}|V\rangle_{2'}+|V\rangle_{1'}|H\rangle_{2'}]|N-1,0\rangle_{\bf 0'} \bigg).
 \end{eqnarray}
If we now introduce the same type of polarization dependent loss as used before to all the modes ${\bf 0'}$, then the weights of the output state are equalized. As a result, the output state becomes
\begin{eqnarray}\label{eq:3e}
 |H\rangle_1|W_{N} \rangle\to \frac{|N,1\rangle}{\sqrt{2N}}=\sqrt{\frac{N+1}{2N}}|W_{N+1}\rangle.
 \end{eqnarray}
Thus, the proposed circuit can expand an input $N$-photon $W$ state into an $(N+1)$ photon $W$ state with the help of an ancilla photon with a success probability of $p(W_N\rightarrow W_{N+1})=1/2+1/2N$, which approaches a constant $1/2$ for large $N$. Starting with an input state $|V\rangle$, the cascaded operation of the circuit $k$ times (together with the polarization dependent losses and required $k$ ancilla photons in the state $|H\rangle$),  one can prepare $(k+1)$ photon $W$ state, $|W_{k+1}\rangle$, with a probability of $p(W_1\rightarrow W_{k+1})=(k+1)2^{-k}$. Finally, we find that cascaded use of this circuit $N$ times, the size of the initial $W$ state $|W_N\rangle$ can be doubled to $|W_{2N}\rangle$ with a success probability of $p(W_N\rightarrow W_{2N})=2^{1-N}$.

We close this section by concluding that the circuit shown in Fig.~2 can deterministically prepare a Bell state, {\it i.e.} $|W_2\rangle$. Preparing $W$ states with this circuit when starting from a single photon state or the state $|W_2\rangle$ requires cascaded operation of the circuit, which prepares the desired state only probabilistically. We should state here that the success probability of this circuit supersedes that of all previously proposed $W$ state preparation and expansion circuits.

\section{Deterministic $W$ state expansion by the parallel use of the circuit}

From Eq.~(\ref{eq:3a}), one can see that although the circuit performs symmetrization
among the input photon and the ancilla photon, the success probability for the $H$-input is twice that for the $V$-input. In other words, the two output terms for the $V$ input have the same amplitude, while the output term for the $H$ input has a different amplitude. This is the reason why the gate performs probabilistically for states having more than 2 photons. Equalizing the weights of all the terms without using polarization dependent loss would lead to deterministic gates. For this purpose, we propose to use the circuit given in Fig.~4 in parallel such that in order to expand an $N$-photon $W$ state $N$ circuits are used, and each of the photons from the input state are input to one circuit together with an ancilla $H$-photon.
\begin{figure}[t]
\centerline{\includegraphics[width=0.6\columnwidth]{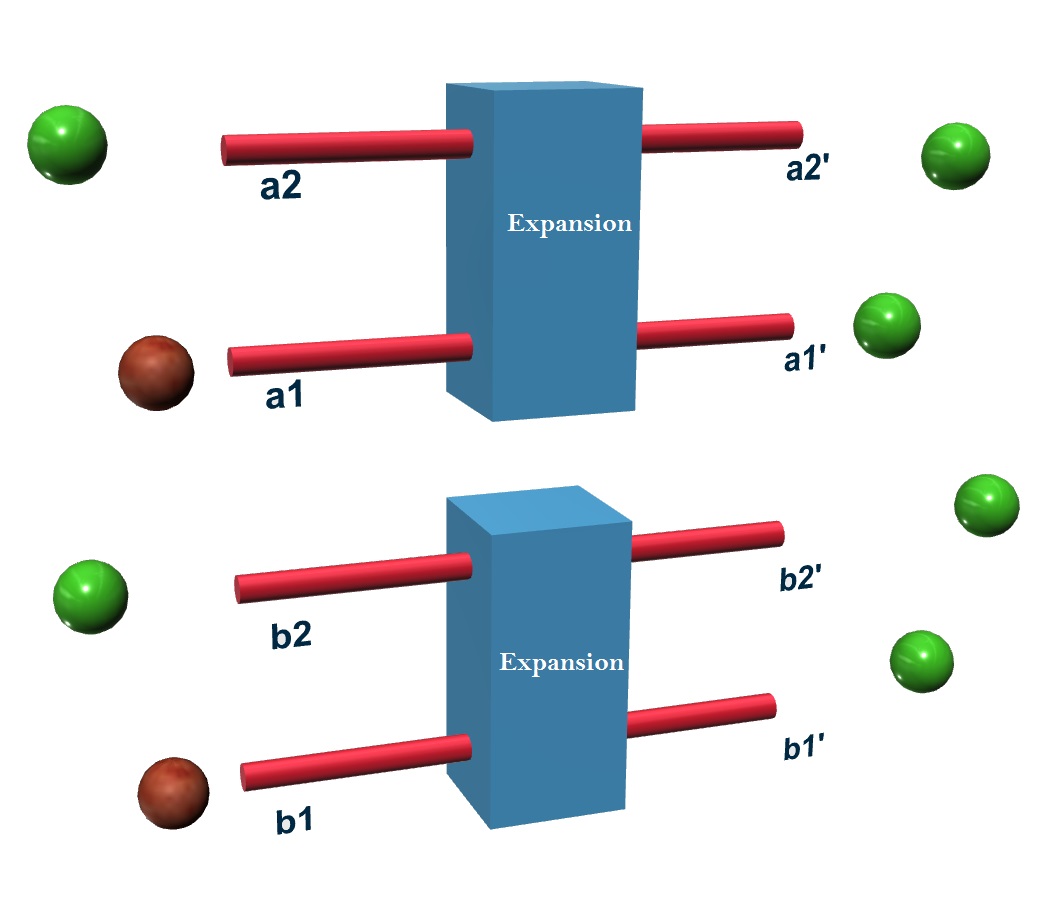}}\label{parallel}
\caption{Deterministic preparation of a $W$ state. Starting from a Bell pair (green) and two independent photons in $|H\rangle$ polarization (red), a $W$ state of four photons (green) can be created deterministically by parallel use of the expansion circuit.}
\end{figure}

Let us consider the input state $|\Psi^{+}\rangle_{a_2b_2}=|W_2\rangle_{a_2b_2}=(|HV\rangle+|VH\rangle)_{a_2b_2}/\sqrt{2}$ and denote the circuits as circuit $a$ and circuit $b$, with their input modes defined as $a_1$ and $a_2$ for the first circuit, and $b_1$ and $b_2$ for the second circuit (see Fig.~4). The photons in modes $a_2$ and $b_2$ are input to the circuit input modes $a_2$ and $b_2$, respectively. The other inputs of the circuits are provided with photons in the state $|H\rangle$. In this case, the total state space of the input can be written as $(|HH\rangle_{a_1a_2}|HV\rangle_{b_1b_2}+|HV\rangle_{a_1a_2}|HH\rangle_{b_1b_2})/\sqrt{2}$. Then the parallel operation of the two circuits on the two modes of the input Bell state leads to
\begin{eqnarray}\label{eq:4a}
 && \hskip-0.5cm |HH\rangle_{a_1a_2}|HV\rangle_{b_1b_2} \to \frac{|HH\rangle_{a_1'a_2'}[|HV\rangle_{b_1'b_2'}+|VH\rangle_{b_1'b_2'}]}{\sqrt{2}} \nonumber \\
 && \\
 && \hskip-0.5cm |HV\rangle_{a_1a_2}|HH\rangle_{b_1b_2} \to \frac{[|HV\rangle_{a_1'a_2'}+|VH\rangle_{a_1'a_2'}]|HH\rangle_{b_1'b_2'}}{\sqrt{2}}. \nonumber \\
 \end{eqnarray}
Clearly not only the symmetrization of the ancilla photons and the input photons is achieved but also the weights of the output terms are all equal. Taking the superposition of the above expressions, we find that the circuit performs the transformation $|\Psi^{+}\rangle=|W_2\rangle\rightarrow|W_4\rangle$, preparing deterministically a four-qubit $W$ state from a two-qubit $W$ state.

Next, let us consider $|W_{3}\rangle$ state for which we showed in the general approach in the previous section that applying the proposed circuit on only one qubit of $|W_{3}\rangle$ prepares $|W_{4}\rangle$ with a probability of $2/3$. Now let us first assume that the first two qubits of the state $|W_{3} \rangle_{123}=\frac{1}{\sqrt{3}}[|HH \rangle_{12} | V \rangle_{3} +\sqrt{2}| W_{2} \rangle_{12} |H \rangle_3]$ are sent simultaneously to two circuits together with the ancilla states. The circuits will perform the following transformations:
\begin{eqnarray}\label{eq:4abc}
 |HH\rangle_{12}|HH\rangle|V\rangle_3&&\to |HHHHV\rangle\\
 |W_2\rangle_{12}|HH\rangle|H\rangle_3&&\to |W_4\rangle|H\rangle,
 \end{eqnarray}
from which we find
$|W_3\rangle\rightarrow \frac{1}{\sqrt{3}}(|HHHH\rangle|V  \rangle+\sqrt{2}|W_4\rangle|H\rangle)$. Rearranging the term and performing a polarization dependent loss on the qubit that does not enter the circuit (qubit 3), we find that the scheme prepares a five-qubit $W$ state $|W_5\rangle$ from the $|W_3\rangle$ with a probability of $5/6$ which is certainly higher than that of the case when only one qubit of $|W_3\rangle$ states enter the circuit. Finally, let us input each of the qubits of the $|W_3\rangle$ in parallel to three circuits together with the ancilla qubits. In this case we have
\begin{eqnarray}\label{eq:4abcd}
 |HHV\rangle_{123}|HHH\rangle&&\to |HHHH\rangle|W_2\rangle\\
 |W_2\rangle_{12}|H\rangle_3|HHH\rangle&&\to |W_4\rangle|HH\rangle
 \end{eqnarray}
which leads to $|W_3\rangle\rightarrow \frac{1}{\sqrt{3}}(|HHHH\rangle|W_2\rangle+\sqrt{2}|W_4\rangle|HH\rangle)=|W_6\rangle$. Thus we conclude that parallel use of three circuits each acting on one qubit of the $|W_3\rangle$ deterministically doubles the size of the $W$ state to $|W_6\rangle$.

It is clear that the balance of the weights of the superposition is destroyed whenever a qubit in the state $|V\rangle$ enters the circuit as this introduces $1/\sqrt{2N}$ as the weight of newly formed two terms. The rest of the terms have the weights $1/\sqrt{N}$. In an $N$-qubit $W$ state $|W_N\rangle$, the qubit in the V-state can be in any of the $N$-possible spatial modes with a probability of $1/N$. Thus, in order to achieve a balanced superposition state each of the $N$ qubits of the $|W_N\rangle$ should be sent to a different circuit together with an ancilla qubit. In this way, all the terms at the final superposition state will have the weight $1/\sqrt{2N}$. Thus, by running $N$-circuits in parallel a $|W_N\rangle$ state can be expanded to $|W_{2N}\rangle$ deterministically.


The new circuit we have proposed and analyzed can double the size of a given $W$ state deterministically. Besides expanding a given $W$ state, the circuit can also create a Bell pair from a qubit in the $V$ state and a four-qubit $W$ state from the Bell pair. We note that the deterministic doubling of the number of qubits in the $W$ state is possible when each of the qubits of the initial $W$ state is input to different circuits running in parallel. In this way, $W$ states with even number of qubits are prepared deterministically. If a $W$ state with an odd number of qubits $W_{2N+1}$ is desired, there are two possible ways to proceed. First, one can prepare the state $W_{2N}$ deterministically and then add one qubit using the same circuit by operating on only one qubit of the $W_{2N}$ state. This can be achieved with a success probability of $1/2+1/4N$. Second, one can prepare the state $W_{2N+2}$ deterministically and then perform a projective measurement on one qubit of this state. If the outcome is in the state $|H\rangle$, {\it i.e.} successful, the remaining $2N+1$ qubits will be in the state $W_{2N+1}$. If the outcome is $|V\rangle$ then all the remaining qubits  will be in the state $|H\rangle$ and entanglement will be destroyed. The probability of the successful outcome is $1-1/2(N+1)$ which approaches 1 for large $N$. The second approach is better for preparing large-scale $W$ states with an odd number of qubits. If all the qubits are not input to circuits operating in parallel, then the preparation and expansion process is probabilistic and the probability of success depends on the number of qubits being operated by the circuits. Even in this case, the probability of success of this circuit is much higher than all of the previously reported $W$ state preparation and expansion gates or circuits. Thus, by repeatedly running the circuit one can start with a $V$-state and prepare $W$ states with a desired number of qubits probabilistically by adding one qubit at a time.

When the number of particles increases beyond two, not only the preparation but also the verification of the multipartite entangled states via techniques such as state tomography become difficult. Regarding $W$ states, however, a recent work showed that a robust self testing of three particles is possible, which can be directly extended to an arbitrary size~\cite{Scarani2014SelfTesting}. The scheme proposed in this work is also capable of testing whether a given state is a genuine $W$ state or not. It can do this by propagating the final state back through the scheme and measuring the output qubits.


\section{Conclusions}

In summary we have proposed an expansion circuit and different schemes based on this circuit to prepare and expand $W$ states. The scheme expands an $N$-qubit $W$ state to a $2N$-qubit $W$ state deterministically with the help of ancilla states by accessing locally all $N$-qubits of the input $W$ state. The circuit consists of a controlled-NOT gate and a controlled-Hadamard gate, which can be decomposed into two controlled-NOT gates and four single-qubit gates. While an all-optical set up was proposed, the scheme could also be applied to various other physical systems such as ion traps, cavity-QED and solid state systems. Compared to previously proposed probabilistic schemes~\cite{Tashima2008,Tashima2009A,Tashima2009B,Tashima2010,Ikuta2011,Ozdemir2011,Bugu2013A,Yesilyurt2013A,Ozaydin2014A}, our scheme is fully deterministic in principle and its success probability is limited only by experimental imperfections. Hadamard and $F$ gates can be implemented efficiently in linear optics using wave plates and a controlled-NOT gate can be implemented as outlined in Ref.~\cite{Brien2003,Nemoto2004,Brien2009,Okamoto2011}. Advances in nonlinear optics will increase the success probability of realizing a photonic-based controlled-NOT and thus the probability of the proposed set up when using polarization-encoded photons.


\section{Acknowledgements}

This work has been funded by Isik University Scientific Research Funding Agency under Grant Number: BAP-14A101. FO thanks to F. Ilhan for fruitful discussions. MST is supported by the South African National Research Foundation and the South African National Institute for Theoretical Physics.


\end{document}